# Technology Integration around the Geographic Information: A State of the Art

Rafael Ponce-Medellín[1], Gabriel Gonzalez-Serna[2], Rocío Vargas[3] and Lirio Ruiz[4]

[1,2,3] Centro Nacional de Investigación y Desarrollo Tecnológico, cenidet
Cuernavaca, Morelos, Mexico

[4] Universidad de la Sierra del Sur, UNSIS
Miahuatlan, Oaxaca, Mexico

**Abstract**
One of the elements that have popularized and facilitated the use of geographical information on a variety of computational applications has been the use of Web maps; this has opened new research challenges on different subjects, from locating places and people, the study of social behavior or the analyzing of the hidden structures of the terms used in a natural language query used for locating a place. However, the use of geographic information under technological features is not new, instead it has been part of a development and technological integration process. This paper presents a state of the art review about the application of geographic information under different approaches: its use on location based services, the collaborative user participation on it, its contextual-awareness, its use in the Semantic Web and the challenges of its use in natural langue queries. Finally, a prototype that integrates most of these areas is presented.

## 1. Introduction

As the integration between technology and society grows up, the task of accessing and retrieving information becomes a crucial point for decision-taking activity, caused by the ever growing user's needs. These needs are resumed in a more natural, faster and pertinent information retrieval; these desirable features are considered because while searching for information, the user experience frequently becomes frustrating due to the impossibility of making a precise petition on the required form, also by the immense quantity of retrieved results or by the imprecision of them, in which there are irrelevant results meanwhile the really relevant responses are not found.

There are different fields in which the differentiation of the diversity of users takes a higher importance. That is the case for the searching of places and geo-located services, through a digital map for example. Under this focus, the use of located based services (LBS) let the user access to information for locating geographic places and points of interest (POIs), making this desirable to find and locate the more adequate ones to a user's particular need.

In recent years, it has become common the use of digital maps in hands of Google Maps, Yahoo Maps, Live Maps, among many others. The relevance of the geographic information has risen a great interest on users and development communities, which is manifested in the huge amount of APIs and mashups applications that have been appearing; there are different location based services, offered by mobile communication and services providers, e.g. there are services for route planning, city guides, hotel guides, etc; most of these applications are commercial or academic products that are generally focused on the locatable contextual dimension.

Other researches that consider the geo-located information to place searching includes the COMPASS 2008 project [1] which presents the use of ontologies towards a context-aware personalized information system, the tourist guide based on location from the Cyberguide project [2] which aimed to provide information to tourists based on their position, or the Lancaster City guide [3] in which according to the location and the user's preferences allows a visitor to get interesting information about the points of interest in the region.

It must be considered that the same user, under different circumstances is going to have different goals and purposes. On the field of geo-located services it is important to consider the users' individual characteristics, because each person has different social, cultural and economical features, making his or her information needs implicitly different. This becomes necessary to have context-aware applications that consider the mentioned features in order to avoid the results that are not relevant to a particular user under specific spatial and temporal situations, at querying for locating places and services, such as hotels or restaurants.

However, since the amount of information has been continuously growing, the remaining data relevance problem and the information overload have become more serious. The use of semantics has been proposed to face up those problems [4], mainly through the use of ontologies. According with this, Perry et al. [5] presented an ontology based model that integrates the thematic, spatial and temporal dimensions, using a high level ontology that defines a general set of thematic classes and spatial entities, with associated relations that connect them. Other works for geographic ontology creation includes the development of a Web tool for this task [6], meanwhile [7], [8] and [9] are focused on the use of geographic domain ontologies for service description as a mean to counteract the ambiguity on user's queries.

The previous boards to another topic: the problems of vagueness inside natural language queries. In this respect, various attempts have been focused towards the identification and nature of natural language quantitative and qualitative prepositions used under geographic information domain with the purpose of a better understanding of determine what the user is trying to mean at querying [10].

The present paper gives a brief overview of the previous technologies and topics mentioned above considering its use related with the search of places and services through the use of geographic information. The uses that geographic information has today seem to cover a variety of research areas, each one with its own challenges. However, it can be seen that the developments among them could be integrated, giving to final users more reliable and personalized systems.

The content of the document is as follows: sections 2 to section 7 covers topics of Location Based Services (LBS), map mashups, collaborative geographic systems, context-awareness on LBS environments,  the integration of geographic related information under the Semantic Web focus and the challenges from natural language querying for geographic information. Section 8 presents a prototype that integrates some of the previous topics, and finally conclusions are presented.

## 2. Geographic Information on Computer Maps and LBS

LBS (Location Based Services) are services that based on user's geographic location, can provide relevant information to his or her geographic position. The main purposes that a LBS system includes are the identification, search and verification of services that are nearly to a user's position, fulfilling tasks as identifying a user's location, locate other persons, objects or places, and provide guidance, information or help to find a particular place [11]. The main characteristic from LBS services is that they provide *just in time* information to the users, considering that the information presented must be from a particular domain of interest to the user and that this information be useful and can be used in the geographic area the user is at.

Steiniger [12] presents LBS systems as an infrastructure composed by: users, a communication network, a positioning component (such as a GPS), LBS service provider and data and content providers. The content provider turns to the service provider in order to obtain the geographic data and information needed to answer the user's query.

LBS design can be considered from two different approaches: *generic services,* in which a user explicitly denotes his or her location (i.e. giving street names or zip code) and *locatable services*, in which the location is automatically obtained in a transparent way to the user, using devices instead, as a GPS.

Also, LBS systems can be classified under different focuses, according to the activities they are going to be used for. For example, they are classified considering if they are going to be used for gaming, getting information or paying bills [13], meanwhile in [14] they are classified according the consumers.

There have been developed different formats in order to facilitate the managing of POIs information and users visited geo-routes between computer devices. These formats can be used for the representation of geographic location on Web maps, for example: LOC, GPX and KML, which are based on XML.

The earlier map applications on mobile devices worked the same visualization that the one used on desktops and on Internet; this presented some troubles due to the different situations that the mobile environment required, because mobile applications have had smaller screen displays in comparison to common desktop applications;

also, mobile applications are commonly used on open spaces, making its users' needs different.

These caused that the first difficulties that the LBS systems confronted, fallen on the type of interaction presented between the user and his or her mobile device, the device itself presented limitations caused by its portability; also on the limited visual representation of the earlier mobile devices, due by the low screen resolution, a limited bandwidth (situation which step by step has been counteracted with mobile technology advances such as the 3G network), and also due to the huge diversity of mobile devices, each one showing different functionalities, causing that it was necessary to develop particular device's applications. Actually, the diversity of installed platforms on mobile devices has started to unify, thanks to the apparition of more opened systems, such as Android.

Today, the market of locating services is a commercial sector at growth, supported by big companies, such as Nokia (with its Nokia Maps and other services), Google (with services as Googke SMS and Google Maps for Mobile), AOL (with MapQuest) or Ask (with Ask Mobile GPS), among many others. At the same time, on the Web can be found an immense variety of applications focused on services geo-locating and routes guidance, for example GPS navigation services (e.g. amAzeGPS). Research work includes those that combine distinct technologies in order to give a mayor variety of creative applications, for example GiMoDig [15], pedestrian navigational guides [16] and such ones that combine augmented reality techniques with contents from Wikipedia [17], or those that applies these technologies for gaming, as in TikGames site or in [13].

This relevance on these types of applications has motivated the apparition of contests that motive the development of new and creative uses that take advantage of these technologies. NAVTEQ Global LBS Challenge is an annual international contest, sponsored by cellular phone companies, among others; this contest mainly promotes mobile LBS applications, with notable projects and participants such as Atlas Book (www.networksinmotion.com/products/atlasbook.html), Where (www.ulocate.com), Bones in Motion (bonesinmotion.com/corp/index.html), W-PlanIT (www.w-planit.com), 8motions (www.8motions.com), among many others, including applications that take advantage of the social networks for geo-locating purposes, such as loopt (www.loopt.com), reLive! (http://relive.abmaps.com), Proxpro (www.proxpro.com), etc. Most of these applications are presented as mashups which integrate geographic information.

## 3. Mashups

Actually, the use of Web maps that combine information from different sources and gives a more complete user's experience has become common: the so called map mashup. The mashups are a genre of interactive Web applications that enable to retrieve content from external sources and create new services; they are a stamp of the Web 2.0. Mashups are mainly composed of an API / content provider (i.e. Google Maps, MapQuest, among a variety of Web map providers), the Web site in which the mashup is presented and by the client's Web browser; they involve a diverse set of interrelated technologies, they usually use AJAX and Web protocols as SOAP y REST; some other are also related with the use of semantic Web, ontologíes and the integration of RSS and ATOM feeds [18].

There are different types of mashups, from videos and photos mashups, until those about sales, searches and news. The case of the digital maps is that they have become a multi purpose centralized tool, used for local business, traffic reports, online dating, among other uses that are related to a specified time and place. The collaborative participation plays an important role in the use of map mashups, as it has been shown in projects such as Wikimapia (www.wikimapia.org) for the collaborative description of world POIs; other case is Tagzania (www.tagzania.com), which is based on the social tagging of places. Considering the previous examples, the map mashups can be used in social searching of places and geo-locatable services in Web maps. This has encouraged the apparition of APIs that facilitate the creation of mashups, such as the ones developed by Google or Yahoo's FireEagle API, which can be used for searching of commercial places, such as restaurants or stores (e.g. Yelp.com site); and there is also the inclusion of geo-location components natively inside Web browsers based on W3C Geolocation Specification [19], which can be used to exploit more similar applications and newer ones. Other initiatives includes WikiMiniAtlas (http://meta.wikimedia.org/wiki/WikiMiniAtlas), that includes geo-referential information inside Wikipedia's entrances about geo-locatable places, or the WikiProjekt Georeferenzierung (de.wikipedia.org/wiki/Wikipedia: WikiProjekt_Georeferenzierung/), which uses an alternative based on textual maps. There are also Web 2.0 sites that facilitate the mashup creation to the final users, as MapBuilder (www.mapbuilder.net), meanwhile other more traditional mashups are concerned on the combination of resources from multiple sources, such as GlobalMotion (www.globalmotion.com) or ONGMAP (ongmap.com). Observing this, it can be manifested that there are a lot of information related on geo-location that can be used on Web maps, and the mashups are a great way to use it.

The activity of the users has become an important part in the operation of the mashups; users can participate actively being volunteers that tag their favorite places or by being consumers of these map mashups geo-location services. However, the social collaboration in relation with the geo-locatable information has a bigger transcendence; this has been indicated in the collaborative geographic information systems.

## 4. Collaborative GIS

The development of GIS dates back to 60's, with the apparition of GIS as computational tools for capturing, managing and transforming data of spatial references to be used in planning and decision-taking processes. Other work that followed after these, was trying to provide a support for the integration between groups of people and for the decision-taking based on geo-referential data, studying also the impact of them on its users. The development of these systems continued and took advantages of the Internet era, in which the Web maps has a fundamental situation, as is mentioned in [20]; this evolution has been passed to technological issues and the adoption of open standards and the collaborative use inside the social networks [21].

The use of GIS has shown to impulse the capacity of community groups to recognize their necessities, priorities and knowledge. It is in this way that the collaborative GIS appeared. Collaborative GIS are an integration of theories, tools and technologies focused on structuring the human participation on group processes of geospatial-decision. Their objective is to use the argumentation, deliberation, the clear structure of maps and the reconciliation between different groups with representative interests [20].

Meanwhile the Web distribution allows freeing of the spatial barriers in the use of GIS, the use of mobile devices helps in the task of recollecting local data. This integration has helped towards a mobility that allows reaching people and places which under other circumstances could not be considered nor contributed to the spatial information of a GIS. The use of GIS interfaces in the form of thin clients over the Internet has made possible to create maps and the download of spatial information without the need of having to install a complete GIS application. This has changed and made easier the way in which the users access and participate in collaborative planning forums; thanks to this, the users of collaborative maps not only receive geographic information, but also let them actively contribute in the process, generating a cycle of mutual feedback between the users and the map service.

## 5. Contextual Information

A context-aware system is a system that considers the context in which is immersed, in order to provide relevant information and services to a user, in which the relevance is defined according the user's task and according to his or her spatial and temporal situation.

There are different senses to define what the context is. For example, on one hand, the context is considered to be all the things that allow to determine a possible set of responses, or on the other hand, context is all the things that are necessary to understand a query; some dictionaries define it as a physical or situational environment, should be political, historical, cultural or from other nature, in which a fact is considered. Nivala [22] considers that context is any information that can be used to characterize the situation o fan entity (person, place or object). In his way, Sun [23] indicates that the context includes from *external aspects* from the environment of a person (i.e. geographic physical features, cultural events) to *interior aspects* (personal fitness and healthy).

In this way, the context can be considered as a pervasive element, which has been researched under different focuses of use, e.g. the context modeling for geographical applications [24], for intelligent personalized touristic guides [23] and also for places querying and searching, taking advantage on the use of ontologies [25], [26], [27].

In an application, the context-awareness can be *active* (this is when the application can be adapted automatically to the context in which it is in, this information can be received actively when the user introduces it directly to the application) and can be *passive* (when the application automatically detects the context, giving to the user the decision of accepting or rejecting the options showed by the application).

The same information can be accessed by the users for different purposes and under different contexts; depending on the user, it will be the context of the application. It is for this that ways of identifying different contexts of use for the same information has been searched.

Among the different contexts identified, Nivala shows a classification of types of contexts, based on previous work of Chen & Kotz y Schilit [22]:
**Computational context.** It considers network connectivity, nearest computer hardware such as printers, etc.

**User context**. It considers user profile, user location, nearest people, actual social situation.
**Physical context**. It considers light, noise, traffic, temperature.
**Temporal context**. It considers the hour, the day of the week, month and year.

Nivala also presents another classification related to types of context, obtained through experimentation with real users. The identified contexts related to mobile map applications were:
**Context: Localization**. This is defined by the current user location.
**Context: System.** Hardware features in which a mobile map is used.
**Context: Use purpose.** It refers to the use that the map is going to have (e.g. purposes for topographic or tourism activities, etc.).
**Context: Time.** It depends on the moment of the day and the season the user is in.
**Context: Physical surrounds.** Surroundings physical and topological features.
**Context: Navigational history.** The blog of use of the user across different geographical areas.
**Context: Orientation.** It refers to the point the user is viewing at.
**Context: Social and cultural features.** It includes cultural specific features of each region, i.e. the particular symbology that is used on certain region, the format of dates used, the weight system used, currency, etc.
**Context: User.** Personal characteristics, as genre, habits, etc.

Another identification of types of contexts is presented in [28], which present broader categories in comparison with the previous one presented:
**Environmental context.** The time and the weather of the user's location.
**User profile.** Preferences and tastes from the user that uses the service.
**Data profile.** It refers to the information related to the geo-locatable services, such as restaurants, hotels, etc; this data is provided by services providers and includes information about schedules, cuisine offered in a restaurant, etc.

The relevance about contextual information on LBS systems relays on the possibility of provide more personalized and elaborated information related to locatable places and services. Inside the LBS systems, the user location in the main component of contextual information and it has the peculiarity that the context in a mobile environment is dynamic. This dynamism causes that the system needs to be constantly adapting, because the user can frequently change his or her physical location, changing with this the features of his or her surroundings.

Considering the previous, it can be concluded that LBS systems can provide more useful information if they considered the user's profile and other contextual information, in order to obtain more precise and personalized results. This involves a necessity of more sophisticated techniques for the mobile market, which includes a more known and detailed user's profile, with his or her historical, preferences and tastes, which could be matched with the offers that are available from the locatable vendors.

Although adding contextual-awareness to a mobile map application can improve the usability of a map [22], it is still necessary to have more intelligent applications which take advantage of the contextual information, making present the challenge of verifying the semantic relevance of the contextual data. One approach for this is the use of Semantic Web technologies.

## 6. Semantic Web

The Semantic Web has been presented as an extension of the current Web. It is part of Tim Berners-Lee's vision about the future of the Web; this vision includes having relationships between different information resources, using additional metadata that enable machines to understand and process the information on the Web; however, the traditional Web does not presents the technology to capture the different relationships concerned with the resources.

A bigger step related to the technological evolution was the appearance of XML, which is one of the basic layers of the Semantic Web. However, XML alone is insufficient, because it only provides a syntactic interoperability. Sharing an XML file adds meaning to its content, but only if the publisher and the receiver can identify and understand the named elements that are in it. This makes necessary research to make more intelligent both data and process modeling.

The way in which data is presented to be processed by machines, relays on making the data more intelligent; for this purpose there have been developed languages and technologies just as:
- *RDF (Resource Description Framework)*. This language is used for describing resources (which can be classes or concepts) and its relationships, in the way of triplets. Triplets are composed by three parts: *subject* (the resource that is being described, and it is identified by an URI), *predicate* (the relation existing between a subject and an IJCSI International Journal of Computer Science Issues, Vol. 5, 2009                                                                                              22 object) and an *object* (an object is a resource that is referred or a literal value instead).
- *RDFS (RDF Schema)*. This language allows creating new classes and properties using RDF; with this it is possible to conform hierarchies of classes for the classification and description of resources.
- *OWL (Web Ontology Language)*. Language for the managing of ontologies. An ontology is an explicit specification of a shared conceptualization [29].
- *SPARQL (SPARQL Protocol and RDF Query Language)*. It is a language for querying semantic Web documents.
- *Rule languages*. Rule languages and formats that allow to infer information from other data. A rule specifies an action that is accomplished if certain conditions are fulfilled (*if (x) then y*). Some examples are SWRL (Semantic Web Rule Language) and W3C's RIF (Rule Interchange Format).

Ontologies allow describing and representing in a computer usable way a portion of a mental model about specific domains; they can be used to achieve a major interoperability among different data sources. The use of ontologies on GIS developments allow the interchange of knowledge and the integration of information; for this purpose, ontologies has been used to define common shared vocabularies, for metadata modeling, for defining the meaning of the data across different domains, for data integration, for classifying resources, for information retrieval, among other uses.

The use of ontologies presents a number of advantages related to the integration of geographic information on the Web, such as:
- They allow making queries based on semantic values.
- They provide the availability of having the information represented under different levels of detail.
- They provide a dynamic access to the information.

Some work that denote the relevance of the use of ontologies for managing geographic information is presented in [30], [31]; the ontologies under the geographic information domain has been worked under different approaches, from Web tools for the creation of geographical information domain [32], to projects that consider the description of GIS services and its use for a matchmaking between different ones in order to allow more complex queries [33], [34]. Fonseca [35] considers that if the use of ontologies is part of an active geographic information system, they conform an *Ontology-Driven Geographic Information Systems* (ODGIS), which presents the advantages of having multiple interpretations (roles) of a same geographic feature; this allows to attend different market sectors, for example, considering a concept as *lake*, it can be used under different circumstances by different geographic information communities: for a department of water studies it is going to be a source of pure water, for a environmental scientist is a wild-life habitat, for a tourist department it is a recreational place, and so on [36].

In order to facilitate the user creation and edition of ontologies, tools as Protégé has appeared; however, there have been appeared new ontology editors that also consider in an innate way the inclusion of georeferential data through Web maps, such as TopBraid Composer.

It is common to find distinct ontologies about the same domain, under a context if not equal, similar. This causes that the use of GIS ontologies shows difficulties due to the peculiarities of this domain, such as the ambiguity of the concepts and the dependent nature on the interpretation and geographical representation of its context. This situation has originated the necessity of establishing a correspondence across the concepts and the relationships between those ontologies. For this purpose, it has been used techniques of matchmaking [37], [38], such as the Normalized Google Distance (NDG) [39].

The selection of the more appropriate vocabularies is a challenge in terms of interoperability; it is because this, that in [40] is recommended the use of terms that come from controlled vocabularies in the form of keyword lists, taxonomies or thesaurus, which can be used to match the data and metadata used in a GIS, improving the quality of the query results.

## 7. Natural Language for Geo-querying

Human communication is flexible, contextualized and full of vagueness and ambiguity. Humans speak and think about spatial relations in an imprecise way, using vague and probabilistic concepts; the people queries and orients in a qualitative mean, not in a quantitative mean [10], meanwhile computers use languages with a well defined semantic and grammar for their communication.

The use of spatial relations in natural language gives to GIS users more alternatives to formulate their queries, according to their task in which they are using a system. The qualitative models can contribute to a broader user of GIS technology.

The real world entities that are represented on a geographical database are corresponded to dots, lines and areas. Generally, geospatial queries include a symbolic representation of the spatial relations, instead of a detailed geometric description; some of these spatial relation models in natural language are present on [10] and [41]:

**Linguistic models.** They are based on introspection, but they present a lack of mathematical objectivity which difficult its use on an information system.
**Geometric models**. These models are oriented towards a quantitative spatial location.
**Connectionistic approaches**. Model has have its basis on the human brain to do parallel computing; it involves training a model using a set of detailed geometric configurations.
**9-intersection model**. This is a topological model, which can be based on cardinal directions or approximated distances.

Based on GIS literature, there are different types of common spatial relations, such as:
**Topological relations**. These are spatial relations that do not vary under continuous transformations.
**Metric relations**. These refer to those relations about distance, which can be quantitative if the distance can be sized or qualitative if the distance is described as near o far.
**Relations of direction**. These makes reference to azimuth, which is quantitative; it also refer to the description of directions in a way such as forward, to the right or to the west, which are qualitative directions.

In order to create formalisms about natural language spatial relations, it is necessary to do a matching between the symbolic representations of the real world spatial relationships, towards a valid geometric configuration.

Natural language has a limited set of words that can be used to describe an infinite number of geometric configurations that can be built. For example, in despite the geometry of a set of configurations which show the path of a way over a park (symbolized has a line and a polygonal shape, respectively), it is possible to have different paths that are different but people would use the same terms to describe these different configurations (i.e. a path across the park, but inside the park there could be different paths that cross it). On the other hand, different terms can be used to describe the same configuration, also if the topology of these configurations is the same (i.e. different people can describe the same path using different terms and expressions).

The use of prepositions in order to define spatial relations is subjected to complex and hidden rules [42]; the spatial relations defined by a natural language are context dependant, i.e. the meaning of *near* is imprecise and depends on the scale that is used, on factors as the geographic region, the moment and place, the task that is performed as well as personal characteristics [43].

Querying for geographical information of places and services can become a difficult task because the implicit problems of having to express which is the desired information to the system. It is because this that different means of querying has been searched, not depending only on a standard textual entrance. It is in this way that the interaction between users and computers relays on the entrances that the firsts give to the machines, letting them express spatial relations with vague terms through natural language queries from mobile devices [44] or through dialogue based systems [45]. Another approach is the use of multi modal interfaces for geographic information related systems; among these, there have been some that have considered spoken entrances and spoken guidance [46], the use of sketches in which a user draws his or her location, such as the ones described on [47] and [48]; other attempts include the use of visual languages [49], [50] and the use of augmented reality technologies [51].

## 8. Case of Study Prototype

A prototype for locating restaurants which considers contextual-awareness and Semantic Web technologies was developed. Data from Chefmoz repository (http://chefmoz.org/) was used; this data is maintained by the social collaborative effort of people across the world. The prototype considers the management of users' and geo-located services' profiles through ontologies for a more pertinent and personalized information retrieval of geo-located places, matching user and services features, in other words, for this case, the goal was matching personal profiles and Chefmoz restaurants' data, considering the personal, spatial and temporal contextual domains of a query.

A process of reverse geocoding over each restaurant described on Chefmoz data was applied, using Google Maps' geocodification functions. This was because Chefmoz data includes the address of each restaurant in it, but it doesn't have geospatial information about latitude and longitude coordinates, which are used to locate a point over a Web map.

An ontological vocabulary focused on representing the user's preferences, including user's characteristics and cuisine preferences was applied as an extension inside the FOAF vocabulary; FOAF is a well known vocabulary that allows describing main data of a user and his or her friends. By adding the proposed vocabulary extension, is a way to show the scalability that the Semantic Web allows.

The user location is obtained from a Web map (if this information was not included in the extension of the user's FOAF file). Until this point, the nearest restaurants to a user specified position can be located, but it is also necessary to determine which of these results are the more pertinent considering the user's preferences.

The contextual models taken into account for this approach include:
- *Service model*. Describes restaurants' specific details.
- *User model*. Describes the user profile.
- *Environment model*. Specifies the state of the elements that surround a user, such as date, hour and weather.

The main concepts taken from Chefmoz that could be used for matching the user model and the environment model are: cuisine, alcohol, smoking, dress, recommended dishes, accepts (type of payment), parking, hours, accessibility and price; these concepts and its respective relationships are used for a restaurant domain ontology. The resulting restaurant ontology is formed from Chefmoz concepts and other domain ontologies found through Swoogle search machine. This ontology is used to represent the service model.

Then, the user model is matched with the restaurant data, using SPARQL queries, in order to find the more accurate restaurants that match with the spatial and personal context of a user, in other words, a matching between the Chefmoz data and the FOAF user file is done.

Until the previous steps, a set of restaurants have been retrieved considering the location and some user's preferences; however, in order to obtain more accurate results, the use of semantic rules can be applied. With these, other contextual information can be matched, according to special situations, e.g. considering the weather (from the environment context) a desirable feature would be expected from the restaurants retrieved (if it is raining, then return restaurants that have under roof characteristic). The idea behind this is also to have a scalable mean for adding and extending more rules; so, according to the values obtained from the context, there are going to be rules that are going to be selected. The language used for applying semantic rules was Semantic Web Rule Language (SWRL) implementation included in Protégé ontology editor; for obtaining a set of results after applying the rules, Semantic Query Web Rule Language (SQWRL) was also used.

For the application of the rules it was necessary to retrieve environment information. For the environment model some Web services that provide information about the weather and the current hour of a particular place have been used. This data is obtained from the latitude and longitude in which user is in; in this way it could be possible to provide relevant results according to the moment a query is made, considering for example the opening and closing hours of a restaurant. Another characteristic related with the personal context is the role that a person is having at the moment of the query, roles as if the person is with his or her family, if he or she is in a business trip and so on.

Finally, FreeLing (a part-of-speech tagger) was used for receiving the input entrance in natural language; after receinving the input, FreeLing identifies the nouns and verbs of the entrance, and using a mediator ontology with a controlled vocabulary (which considers the terms presented on Chefmoz) the query in matched with the restaurant ontology. With this process is possible to determine the information that is desired from the service model.

A resume of the steps explained in this section can be observed on Fig. 1. As future work on this prototype, it is the adding of more rules, the adding of more terms to the controlled vocabulary used for the FreeLing natural language process; finally the prototype is going to be tested with real users.

## 4. Conclusions

The continuous and dynamic changes on the technology adoption by the society have turned the different technological uses of geographic information and applications a crucial component for the decision-taking, inclusive on particular issues, such as the election of the nearest restaurants that provides the most according features against the user preferences.

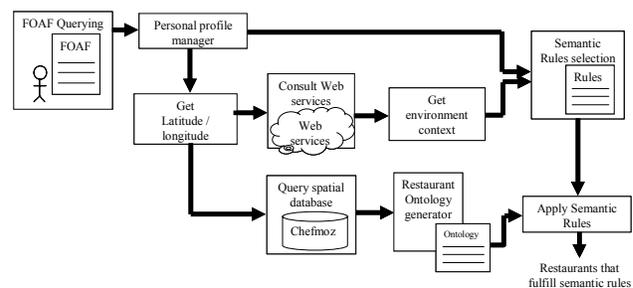

Fig. 1 Steps considered for a prototype that integrates context-awareness, Semantic Web technologies and data obtained from Chefmoz collaborative project, in order to locate the more accurate and personalized selection of restaurants for a user.

The combination of different data sets, for example through mashups, coincides with the essence of the Semantic Web; however, the Semantic Web provides of more consistent models and tools used to define and use the implicit relationships among the data on the Web, and as well as the Web 2.0, a more intelligent share of information is desired.

Thanks to the use of social networks and collaborative participation have been appeared geo-social networks that interact with digital map services and organize the social collaboration among the users and their geographical locations, an example of this is the Chefmoz project.

The availability of semantically annotated data is crucial, such as the Chefmoz project; it is possible to apply a collaborative perspective as a data source of information. However, in our practice, although Chefmoz has a variety of annotated information, it presents several bugs, making its processing a slowly but no impossible task.

The consideration of semantically contextual data can be used in order to improve the precision of the retrieved results for geo-location queries. It can be seen that the next generation of geospatial applications should combine real world knowledge with space and time context, bringing more precise and personalized geo-location retrieval systems.

## References

[1] B. Holtkamp, N. Weißenberg, M. Wojciechowski, R. Gartmann, "Matching Dynamic Demands of Mobile Users with Dynamic Service Offers", Handbook of Ontologies for Business Interaction, IGI Global, 2008, pp. 278 – 293.
[2] S. Long, R. Kooper, G. D. Abowd, C. G. Atkeson, "Rapid prototyping of mobile context-aware applications: The cyberguide case study", 2nd ACM International Conference on Mobile Computing and Networking, ACM, 1996, pp. 97 – 107.
[3] N. Davies, K. Mitchell, K. Cheverest, G. Blair, "Developing a context sensitive tourist guide", First Workshop on Human Computer Interaction with Mobile Devices, GIST Tech. Rep. G98-1, 1998, pps. 17-24.
[4] T. Berners-Lee, J. Hendler, O. Lassila, "The Semantic Web", Sci., Am 284 (5), 2001, pps. 35-43.
[5] M. Perry, A. Sheth, I. B. Arpinar, F. Hakimpour, "Geospatial and Temporal Semantic Analytics", Handbook of Research on Geoinformatics, IGI Global, 2009, pps. 161-170.
[6] E. Tomai, M. Spanaki, "From ontology design to ontology implementation: A web tool for building geographic ontologies", 8th AGILE Conference on Geographic Information Science, 2006.
[7] L. Bernard, U. Einspanier, M. Lutz, C. Portele, "Interoperability in GI Service Chains – The Way Forward", Proceedings of the 6th AGILE, 2003.
[8] E. Klien, M. Lutz, W. Kuhn, "Ontology-based discovery of geographic information services – An application in disaster management", Computers, Environment and Urban Systems, Elsevier, 2005.
[9] M. Lutz, E. Klien, "Ontology-based Retrieval of Geographic Information", International Journal of Geographic Information Science, 2006.
[10] Egenhofer M., Shariff R., Metric Details for Natural-Language Spatial Relations, ACM Transactions on Information Systems 16 (4), EUA, 1998.
[11] Sarjakoski T., Sarjakoski L. T., The GiMoDig public final report, Geospatial info-mobility service by real-time data integration and generalization, http://gimodig.fgi.fi.
[12] Steiniger S., Neun M., Alistair E., Foundations of Location Based Services, Zurich University, Switzerland. 2006.
[13] Lonthoff J., Ortner E., Mobile Location-Based Gaming as Driver for Location-Based Services (LBS) – Exemplified by Mobile Hunters, Informatica Journal, Vol. 31, Num. 2, 2007.
[14] Wehn de Montalvo, U., Ballon, P., van de Kar, E., Maitland, C. Business models for location-based services, AGILE Conference, Salt Lake City, 2003.
[15] Hampe M., Sester M., Harrie L., Multiple Representation Databases to support Visualization on Mobile Devices, Geoinformation Science Journal, Vol. 6, Nr. 1, 2006.
[16] Birgit E., Pedestrian Navigation – Creating a tailored geodatabase for routing, 4th Workshop on Positioning, Navigation and Communication, WPNC '07, 2007.
[17] Schoning J., Rohs M., Kruger A., Paper Maps as an Entry Point for Tourist to Explore Wikipedia Content, Munster University, Germany, 2007.
[18] Merrill D., Mashups: The new breed of Web app, http://www.ibm.com/developerworks/xml/library/x-mashups.html, IBM, 2006.
[19] Popescu A., Geolocation API Specification, W3C, http://dev.w3.org/geo/api/spec-source.html.
[20] Balram S., Dragicévic S., Collaborative Geographic Information Systems: Origins, Boundaries and Structures, Collaborative Geographic Information Systems, Idea Group Publishing, pps. 1-18, 2006.
[21] Mac Gillavry E., Collaborative Mapping and GIS: An Alternative Geographic Information Framework, Collaborative Geographic Information Systems, Idea Group Publishing, pps. 103-119, 2006.
[22] Nivala A., Sarjakoski L. T., An Approach to Intelligent Maps: Context Awareness, Workshop HCI in mobile Guides, Italy, 2003.
[23] Sun Y., Lee L., Agent-based Personalized Tourist Route Advice System, Geo-Imagery Brinding Continents, XXth ISPRS Congress, Turkey, 2004.
[24] Spaccapietra S., MADS: SPACE, TIME AND CONTEXT MODELING FOR GEOGRAPHIC APPLICATIONS, 3rd eBZ Workshop on e-Government - Organizing the City: Time, Space, and Mobility, Italy, 2006.
[25] Spaccapietra S., WHAT WOULD YOU LIKE TO DRINK?, Context Representation and Reasoning Workshop, ECAI 2006 Conference, 2006.
[26] Yu S., Spaccapietra S., Cullot N., Aufaure M., User Profiles in Location-based Services: Make Humans More Nomadic and Personalized, Swiss Federal Institute of Technology, Switzerland, 2004.
[27] Benslimane D., Vangenot C., Roussey C., Arara A., Multirepresentation in ontologies, Universidad Lyon, France, 2003.

[28] Yu S., Al-Jadir L., Spaccapietra S., Matching User's Semantics with Data Semantics in Location-Based Services, Swiss Federal Institute of Technology, Switzerland, 2005.

[29] Gruber T. R., A translation approach to portable ontologies. Knowledge Acquisition, 5(2):199-220, 1993

[30] Spaccapietra S., On spatial ontologies, GeoInfo 2004, Swiss Federal Institute of Technology, Switzerland, 2004.

[31] Pfoser D., Tryfona N., Data Semantics in Location-based Services, Research Academic Computer Technology Institute, Greece, 2005.

[32] Tomai E., Spanaki M., From ontology design to ontology implementation: A web tool for building geographic ontologies, 8th AGILE Conference on Geographic Information Science, Portugal, 2006.

[33] Lutz M., Ontology-based Discovery and Composition of Geographic Information Services, PhD. thesis, Munster University, Germany, 2005.

[34] Lutz M., Klien E., Ontology-based Retrieval of Geographic Information, International Journal of Geographic Information Science, 2006.

[35] Fonseca F., Ontology-Driven Geographic Information Systems, PhD thesis, Maine University, USA, 2001.

[36] Fonseca F., Egenhofer M., Agouris P. Camara C., Using Ontologies for Integrated Geographic Information Systems, Transactions in GIS 6(3), Maine University, USA, 2002.

[37] Aleksovski Z., Klein M., Kate W., Harmelen F., Matching Unstructured Vocabularies using a Background Ontology, Vrije Universiteit, Netherlands, 2006.

[38] Anaby-Tavor A., Gal A., Trombetta A., Evaluating Matching Algorithms: the Monotonicity Principle, Israel Institute of Technology, 2003.

[39] Gligorov R., Aleksovski Z., Kate W., Harmelen F., Using Google Distance to weight approximate ontology matches, International World Wide Web Conference Committee (IW3C2), 2007.

[40] Lacasta J., Nogueras-Iso J., Béjar R., Muro-Medrano P.R., Zarazaga-Soria F.J., A Web Ontology Service to facilitate interoperability within a Spatial Data Infrastructure: applicability to discovery, Data & Knowledge Engineering Volume 63 (3), Elsevier Science Publishers B. V., 2007.

[41] Shariff R., Talking GIS: A Theoretical Basis, Asian Association on Remoting Sensing, ACRS 1997.

[42] Spatial Cognition and GIS Customisation, Gisca, http://www.gisca.adelaide.edu.au/education_training/lectures/sdvis5014/lectures/lecture13.html.

[43] Nogueras F., Latre M., Navas M., Rioja R., Muro P., Towards the construction of the Spanish National Geographic Information Infrastructure, EC-GIS 2001, 7th European Commission GI & GIS Workshop, Managing the Mosaic, 2001.

[44] Le X., Yang C., Yu W., Chen F., Natural Language Query Interface in SMS/MMS-based Spatial Information Service, Geoscience and Remote Sensing Symposium, IGARSS '05. Proceedings, IEEE International, Chinese Academy of Sciences, China, 2005.

[45] Vanderheiden, G., Zimmerman, G., Blaedow, K., Trewin, S. Hello, What Do You Do? Natural Language Interaction with Intelligent Environments, Proceedings of the 2005 HCII Las Vegas Conference, 2005.

[46] Cai G., Contextualization of Geospatial Database Semantics for Human-GIS Interaction, Geoinformatic, USA, 2007.

[47] Cai G., Wang H., MacEachren A., Fuhrmann S., Natural Conversational Interfaces to Geospatial Databases, Transactions in GIS 9(2), 2005..

[48] Caduff David, Egenhofer M., Geo-Mobile Query-by-Sketch, Int. Journal of Web Engineering and Technology, Zurich University, Switzerland, 2007.

[49] Loranca M., Consultas espaciales en una arquitectura de componentes GIS, Tesis de maestría, Universidad de las Américas, Mexico, 2000.

[50] Pacheco A., GEOSIG: Generación de consultas en un Sistema de Información Geográfica, Universidad de las Américas, Mexico, 2002.

[51] Bobrich J., Otto S., Augmented Maps, International Archives of Photogrammetry Remote Sensing and Spatial Information Sciences Vol. 34, part 4, England, 2002.